\documentclass[12pt,thmsa]{article}
\usepackage{sw20lart}

\input{tcilatex}
\begin{document}

\title{Annihilation Contributions in $B\rightarrow K_{1}\gamma $ decay in
next-to-leading order in LEET and CP-asymmetry}
\author{M. Jamil\ Aslam \\
National Centre for Physics and Department of Physics,\\
Quaid-i-Azam University, \\
Islamabad, Pakistan}
\maketitle

\begin{abstract}
The effect of weak annihilation and $u$-quark penguin contribution on the
branching ratio $B\rightarrow K_{1}\gamma $ at next-to-leading order of $%
\alpha _{s}$ are calculated using LEET approach. It is shown that the value
of LEET form factor remains the same in the range of unitarity triangle
phase $\alpha $ favored by the Standard Model. CP-asymmetry for above
mentioned decay has been calculated and its suppression due to the hard
spectator correction has also been incorporated. In addition, the
sensitivity of the CP-asymmetry on the underlying parameters has been
discussed.
\end{abstract}

Exclusive decays involving $b\rightarrow s\gamma $ transition are best
exemplified by the decay $B\rightarrow K^{*}\gamma $, which provide abundant
issue for both theorists and experimentalists. Higher resonances of kaons
such as $K_{2}^{*}\left( 1430\right) $ are also measured by CLEO\cite
{CLEOetal} and the B factories \cite{Bfactory1, Bfactory2}. Recently, Belle 
\cite{belle4led} has announced the first measurement of $B\rightarrow
K_{1}^{+}(1270)\gamma $ 
\begin{equation}
\mathcal{B}(B^{+}\rightarrow K_{1}^{+}\gamma )=(4.28\pm 0.94\pm 0.43)\times
10^{-5}  \label{01a}
\end{equation}
There are several reason to focus on higher kaon resonances. First and the
most promising is that they share lot of things with $B\rightarrow
K^{*}\gamma $, like at quark level both of them are governed by $%
b\rightarrow s\gamma $. Therefore all the achievements of $b\rightarrow
s\gamma $ can be used in these decays, e.g. the same operators in the
operator product expansion and the same Wilson coefficients that are
available. The light cone distribution amplitudes (DA) are same except the
overall factor of $\gamma _{5}$ and this gives few differences in many
calculations\cite{Lee}. Secondly, it was suggested that $B\rightarrow K_{%
\text{res}}\left( \rightarrow K\pi \pi \right) \gamma $ can provide a direct
measurement of the photon polarization\cite{Gronau} and it was shown that
large polarization asymmetry $\approx 33\%$ has been produced due to decay
of $B$ meson through the kaon resonances. In the presences of anomalous
right-handed couplings, the polarization can be severely reduced in the
parameter space allowed by current experimental bounds of $B\rightarrow
X_{s}\gamma $. It was also argued that the $B$ factories can now make a lot
of $B\bar{B}$ pairs, enough to check the anomalous couplings through the
measurement of the photon polarization.

The theorists are also facing challenges from the discrepancy between their
predictions and experiments. It was pointed out that the form factor
obtained using the LEET approach for $B\rightarrow K^{*}\gamma $ is found to
be smaller compared to the values obtained by QCD sum rules or light-cone
sum rules (LCSR)\cite{Ali}. At this stage, the source of this mismatch is
not well understood.

On $B\rightarrow K_{1}\gamma $ side the situation is more complicated. Based
on the QCDF framework combined with the LCSR results, it is predicted that $%
\mathcal{B}(B^{0}\rightarrow K_{1}^{0}(1270)\gamma )=(0.828\pm 0.335)\times
10^{-5}$ at the NLO of $\alpha _{s}$ which is very small as compared to the
experimental value [cf. Eq. (\ref{01a})] \cite{Lee}. The value of the
relevant form factor has been extracted from the experimental data and its
value is found to be $F_{+}^{K_{1}(1270)}(0)=0.32\pm 0.03$ which is very
large as compared to $F_{+}^{K_{1}(1270)}(0)|_{\mathrm{LCSR}}=0.14\pm 0.03$
obtained by the LCSR. This is contrary to the case of $B\rightarrow
K^{*}\gamma $ where the form factor obtained from LCSR is larger than the
LEET\ one and the source of discrepancy is not yet known. But for $%
B\rightarrow K_{1}\gamma $ case the possible candidates to explain this
discrepancy, like higher twist effects in DA, non zero mass effects of axial
kaon, the framework of QCDF, possible mixing in $K_{1}\left( 1270\right) $
and $K_{1}\left( 1400\right) $ and annihilation topologies, have also been
discussed in detail in the literature\cite{Leenew}. The calculation done in 
\cite{Leenew} is for the leading twist and it was pointed out that higher
twist may have some effect on the form factors because all others are not
the suitable candidates. Recently it has been shown that the value of form
factor is not sensitive to the higher twists \cite{Jamil}.

In this paper the effect of weak annihilation and also the $u$-quark
contribution $A^{u}$ from the penguin to the branching ratio for $%
B\rightarrow K_{1}\gamma $ at NLO of $\alpha _{s}$ are calculated using the
LEET\ approach \cite{Dugan,Charles}. We have followed the same frame work as
done by Ali et. al.\cite{Ali} for $B\rightarrow K^{*}\gamma $, because $%
B\rightarrow K_{1}\gamma $ shares many things with it. As it is pointed out
in the literature on $B\rightarrow K^{*}\gamma $, the effect of annihilation
contribution to the charmed quark part of amplitude is numerically small
because only the penguin operator with tiny Wilson coefficients can
contribute. On the other hand the annihilation contribution to the up-quark
part of the amplitude contributes significantly because of large Wilson
coefficients but again the CKM suppression $\left| \lambda
_{u}^{(s)}/\lambda _{c}^{(s)}\right| \approx 0.02$ puts this large
correction for $B\rightarrow K_{1}\gamma $ into perspective\cite{Bosch}.
Finally, by incorporating these annihilation and $u$-quark contributions we
compute the CP-asymmetry $\mathcal{A}_{CP}\left( K_{1}^{\pm }\gamma \right) $
involving the decay $B\rightarrow K_{1}\gamma $. The CP-asymmetry arises due
to the interference of the various penguin amplitudes which have clashing
weak phases, with the required strong interaction phase provided by the $%
\mathcal{O}\left( \alpha _{s}\right) $ corrections entering the penguin
amplitudes via the Bander-Silverman-Soni (BSS) mechanism\cite{BSS}. We find
that the hard spectator corrections reduce the CP-asymmetry calculated from
the vertex contributions alone. The resulting CP-asymmetry, depend rather
sensitively on the ratio of the quark masses $m_{c}/m_{b}$. This parametric
dependence, combined with the scale dependence of $\mathcal{A}_{CP}\left(
K_{1}^{\pm }\gamma \right) $ makes the prediction of direct CP-asymmetry
rather unreliable and present work will be devoted to this issue.

The effective Hamiltonian for $b\rightarrow s\gamma $ can be written as 
\begin{equation}
\mathcal{H}_{\mathrm{eff}}(b\rightarrow s\gamma )=-\frac{G_{F}}{\sqrt{2}}%
V_{tb}V_{ts}^{*}\sum_{i=1}^{8}C_{i}(\mu )O_{i}(\mu )~,  \label{2}
\end{equation}
where 
\begin{eqnarray}
O_{1}^{(p)} &=&({\bar{s}}_{i}p_{j})_{V-A}(\bar{p}_{j}b_{i})_{V-A}~, 
\nonumber \\
O_{2}^{(p)} &=&({\bar{s}}_{i}p_{i})_{V-A}(\bar{p}_{j}b_{j})_{V-A}~, 
\nonumber \\
O_{3} &=&({\bar{s}}_{i}b_{i})_{V-A}\sum_{q=u,c,t}({\bar{q}}_{j}q_{j})_{V-A}~,
\nonumber \\
O_{4} &=&({\bar{s}}_{i}b_{j})_{V-A}\sum_{q=u,c,t}({\bar{q}}_{j}q_{i})_{V-A}~,
\nonumber \\
O_{5} &=&({\bar{s}}_{i}b_{i})_{V-A}\sum_{q=u,c,t}({\bar{q}}_{j}q_{j})_{V+A}~,
\nonumber \\
O_{6} &=&({\bar{s}}_{i}b_{j})_{V-A}\sum_{q=u,c,t}({\bar{q}}_{j}q_{i})_{V+A}~,
\nonumber \\
O_{7} &=&\frac{em_{b}}{8\pi ^{2}}{\bar{s}}_{i}\sigma ^{\mu \nu }(1+\gamma
_{5})b_{i}F_{\mu \nu }~,  \nonumber \\
O_{8} &=&\frac{g_{s}m_{b}}{8\pi ^{2}}{\bar{s}}_{i}\sigma ^{\mu \nu
}(1+\gamma _{5})T_{ij}^{a}b_{j}G_{\mu \nu }^{a}~.  \label{02a}
\end{eqnarray}
Here $i,j$ are color indices and $p$ stands for $u$ or $c$ quark. We neglect
the CKM element $V_{ub}V_{us}^{*}$ as well as the $s$-quark mass. The
leading contribution to $B\rightarrow K_{1}\gamma $ comes from the
electromagnetic operator $O_{7}$ as shown in Fig.\ 1.

As in the case of the real photon emission ($q^{2}=0$), the only form factor
appears in the calculation is $\xi _{\bot }^{(K_{1})}$. Therefore one can
write 
\begin{eqnarray}
\langle O_{7}\rangle _{A} &\equiv &\langle K_{1}(p^{\prime },\epsilon
)\gamma (q,e)|O_{7}|B(p)\rangle  \nonumber \\
&=&\frac{em_{b}}{4\pi ^{2}}\xi _{\bot }^{(K_{1})}\left[ \epsilon ^{*}\cdot
q(p+p^{\prime })\cdot e^{*}-\epsilon ^{*}\cdot e^{*}(p^{2}-p^{\prime
2})+i\epsilon _{\mu \nu \alpha \beta }e^{*\mu }\epsilon ^{*\nu }q^{\alpha
}(p+p^{\prime })^{\beta }\right] ,  \nonumber \\
&&  \label{03}
\end{eqnarray}
with $\epsilon ^{*\nu }$ and $e^{\mu }$ being the polarization vector for
axial kaon and the photon respectively. The decay rate is straight forwardly
obtained to be \cite{Lee} 
\begin{equation}
\Gamma (B\rightarrow K_{1}\gamma )=\frac{G_{F}^{2}\alpha m_{b}^{2}m_{B}^{3}}{%
32\pi ^{4}}|V_{tb}V_{ts}^{*}|^{2}\left( 1-\frac{m^{2}}{m_{B}^{2}}\right)
^{3}|\xi _{\bot }^{(K_{1})}|^{2}|C_{7}^{\text{eff(0)}}|^{2}~,  \label{04}
\end{equation}
where $\alpha $ is the fine-structure constant and $C_{7}^{\mathrm{eff(0)}}$
is the effective Wilson coefficient at leading order$.$

At next to leading order of $\alpha _{s}$, one has to consider the
contribution from operator $O_{2}$ and $O_{8}$ along with that of the $O_{7}$
in $B\rightarrow K_{1}\gamma $ decay. For operator $O_{7}$ all the
subleading contributions shown in Fig. 2 are absorbed in the form factor
where as the Wilson coefficient contains next to leading order parts 
\[
C_{7}^{eff}\left( \mu \right) =C_{7}^{eff\left( 0\right) }\left( \mu \right)
+\frac{\alpha _{s}\left( \mu \right) }{4\pi }C_{7}^{eff\left( 1\right)
}\left( \mu \right) 
\]
On the other hand, for operators $O_{2}$ and $O_{8}$ the leading order $%
C_{2}^{(0)}$ and $C_{8}^{(0)}$ are sufficient for $C_{2}$ and $C_{8}$
because these operators contribute at NLO. Each operator has its vertex
contribution and hard spectator contribution terms which are calculated
explicitly in \cite{Jamil} and are depicted in Figs. [2-6]

The branching ratio for $B\rightarrow K_{1}\gamma $ is given by 
\begin{eqnarray}
\mathcal{B}_{\mathrm{th}}(B\to K^{*}\gamma ) &=&\tau _{B}\,\Gamma _{\mathrm{%
th}}(B\to K^{*}\gamma )  \nonumber \\
&=&\tau _{B}\,\frac{G_{F}^{2}\alpha |V_{tb}V_{ts}^{*}|^{2}}{32\pi ^{4}}%
\,m_{b,\mathrm{pole}}^{2}\,M^{3}\,\left[ \xi _{\perp }^{(K_{1})}\right]
^{2}\left( 1-\frac{m_{K^{*}}^{2}}{M^{2}}\right) ^{3}\left| C_{7}^{(0)\mathrm{%
eff}}+A^{(1)}(\mu )\right| ^{2}  \nonumber \\
&&  \label{branching1}
\end{eqnarray}
where~$G_{F}$ is the Fermi coupling constant, $\alpha =\alpha (0)=1/137$ is
the fine-structure constant, $m_{b,\mathrm{pole}}$ is the pole $b$-quark
mass, $M$~and $m_{K_{1}}$ are the $B$- and $K_{1}$-meson masses, and~$\tau
_{B}$ is the lifetime of the~$B^{0}$- or $B^{+}$-meson. The value of these
constants is used from\cite{Ali} for the numerical analysis. For this study,
we consider $\xi _{\perp }^{(K_{1})}$ as a free parameter and we will
extract its value from the current experimental data on $B\to K_{1}\gamma $
decays.

The function~$A^{(1)}$ in Eq.~(\ref{branching1}) can be decomposed into the
following three components: 
\begin{equation}
A^{(1)}(\mu )=A_{C_{7}}^{(1)}(\mu )+A_{\mathrm{ver}}^{(1)}(\mu )+A_{\mathrm{%
sp}}^{(1)K_{1}}(\mu _{\mathrm{sp}})~  \label{eq:A1tb}
\end{equation}
Here, $A_{C_{7}}^{(1)}$ and $A_{\mathrm{ver}}^{(1)}$ are the $O(\alpha _{s})$
(i.e. NLO) corrections due to the Wilson coefficient~$C_{7}^{\mathrm{eff}}$
and in the $b\to s\gamma $ vertex, respectively, and $A_{\mathrm{sp}%
}^{(1)K_{1}}$ is the $\mathcal{O}(\alpha _{s})$ hard-spectator corrections
to the $B\to K_{1}\gamma $ amplitude and their explicit expressions are as
follows: 
\begin{eqnarray}
A_{C_{7}}^{(1)}(\mu ) &=&\frac{\alpha _{s}(\mu )}{4\pi }\,C_{7}^{(1)\mathrm{%
eff}}(\mu ),  \label{eq:A1tb-C7} \\
A_{\mathrm{ver}}^{(1)}(\mu ) &=&\frac{\alpha _{s}(\mu )}{4\pi }\left\{ \frac{%
32}{81}\left[ 13C_{2}^{(0)}(\mu )+27C_{7}^{(0)\mathrm{eff}}(\mu
)-9\,C_{8}^{(0)\mathrm{eff}}(\mu )\right] \ln \frac{m_{b}}{\mu }\right.
\label{eq:A1tb-ver} \\
&-&\left. \frac{20}{3}\,C_{7}^{(0)\mathrm{eff}}(\mu )+\frac{4}{27}\left(
33-2\pi ^{2}+6\pi i\right) C_{8}^{(0)\mathrm{eff}}(\mu
)+r_{2}(z)\,C_{2}^{(0)}(\mu )\right\} ,\qquad  \nonumber \\
A_{\mathrm{sp}}^{(1)K_{1}}(\mu _{\mathrm{sp}}) &=&\frac{\alpha _{s}(\mu _{%
\mathrm{sp}})}{4\pi }\,\frac{2\Delta F_{\perp }^{(K_{1})}(\mu _{\mathrm{sp}})%
}{9\xi _{\perp }^{(K_{1})}}\left\{ 3C_{7}^{(0)\mathrm{eff}}(\mu _{\mathrm{sp}%
})\right.  \label{eq:A1tb-sp} \\
&+&\left. C_{8}^{(0)\mathrm{eff}}(\mu _{\mathrm{sp}})\left[ 1-\frac{%
6a_{\perp 1}^{(K_{1})}(\mu _{\mathrm{sp}})}{\left\langle \bar{u}%
^{-1}\right\rangle _{\perp }^{(K_{1})}(\mu _{\mathrm{sp}})}\right]
+C_{2}^{(0)}(\mu _{\mathrm{sp}})\left[ 1-\frac{h^{(K_{1})}(z,\mu _{\mathrm{sp%
}})}{\left\langle \bar{u}^{-1}\right\rangle _{\perp }^{(K_{1})}(\mu _{%
\mathrm{sp}})}\right] \right\} .  \nonumber
\end{eqnarray}
The terms proportional to $\Delta F_{\perp }^{(\rho )}(\mu _{\mathrm{sp}})$
above are the $O(\alpha _{s})$ hard-spectator corrections which should be
evaluated at the typical scale $\mu _{\mathrm{sp}}=\sqrt{\mu \Lambda _{%
\mathrm{H}}}$ of the gluon virtuality. The complex function~$r_{2}(z)$ of
the parameter $z=m_{c}^{2}/m_{b}^{2}$, and the Wilson coefficients in the
above equations can be found in Refs.~\cite{Greub:1996tg,Chetyrkin:1997vx};
the function~$h^{(\rho )}(z,\mu )$ and the dimensionless quantity $\Delta
F_{\perp }^{(\rho )}(\mu )$ are defined through Eqs.~(25) and~(27),
respectively of Ref.\cite{Jamil}. Now $C_{7}^{(1)\mathrm{eff}}(\mu )$ and $%
A_{\mathrm{ver}}^{(1)}(\mu )$ are process independent and encodes the QCD\
effects only, whereas $A_{\mathrm{sp}}^{(1)}(\mu _{\mathrm{sp}})$ contains
the key information about the out going mesons. The factor $\frac{6a_{\perp
1}^{(K_{1})}(\mu _{\mathrm{sp}})}{\left\langle \bar{u}^{-1}\right\rangle
_{\perp }^{(K_{1})}(\mu _{\mathrm{sp}})}$ appearing in the Eq. (\ref
{eq:A1tb-sp}) is arising due to the Gegenbauer moments.

By calculating the numerical value from the above expressions and varying
the parameters in the standard range, the value of the form factor is
extracted from the experimental measurements (\ref{01a}) and it was found to
be \cite{Leenew} 
\[
\xi _{\perp }^{(K_{1})}(0)=0.32\pm 0.03 
\]
which is for the leading twist and remains unchanged if one includes the
higher twist effects \cite{Jamil}.

It is already pointed out in the literature that it is unlikely if
annihilation topology would give considerable contributions \cite{Leenew},
but these are important if one wants to study the CP-asymmetry and this is
one of the purpose of this article. Before calculating the CP-asymmetry we
will check the effects of annihilation contribution on the branching ratio
of $B\rightarrow K_{1}\gamma $ decays.

Since weak annihilation is the power correction, we will content ourself
with the lowest order result $\left( O\left( \alpha _{s}^{0}\right) \right) $
for our estimate and to check its effect on the branching ratio. The reason
for including this class of power corrections is that they come with
numerical enhancement from the large Wilson coefficients $C_{1\text{,}2}$ $%
\left( C_{1}\approx 3C_{7}\right) $ but are CKM suppressed and thus these
contributions are expected to be very small for the decay under
consideration. The amplitude for charged $B$ meson decay in terms of Weak
annihilation $A$, charmed penguin $P_{c}$, gluonic penguin $M$ and short
distance amplitude $P_{t}$ can be written as [following the notation of \cite
{Pirjol}] 
\begin{eqnarray}
A\left( B^{-}\rightarrow K_{1}^{-}\gamma \right) &=&\lambda _{u}^{\left(
s\right) }a+\lambda _{t}^{\left( s\right) }p  \label{annihi1} \\
A\left( B^{0}\rightarrow K_{1}^{0}\gamma \right) &=&\lambda _{t}^{\left(
s\right) }\left( P_{t}+\left( M^{\left( 1\right) }-P_{c}^{\left( 1\right)
}\right) +\frac{2}{3}\left( M^{\left( 2\right) }-P_{c}^{\left( 2\right)
}\right) \right)  \label{annihi2}
\end{eqnarray}
where $\lambda _{q}^{\left( s\right) }=V_{qb}V_{qs}^{*}$, $a=A-P_{c}$ and $%
p=P_{t}+M-P_{c}$. As it is known \cite{Pirjol} 
\[
P_{c}\simeq 0.2A\text{, }A\simeq 0.3P_{t} 
\]
i.e. we can safely neglect charmed penguin $P_{c}$ and gluonic penguin $M$
amplitudes relative to the short-distance amplitude $P_{t}$ and weak
annihilation amplitude $A.$ Thus Eq. (\ref{annihi1}) becomes 
\begin{eqnarray*}
A\left( B^{-}\rightarrow K_{1}^{-}\gamma \right) &=&\lambda _{t}^{\left(
s\right) }p\left( 1+\frac{\lambda _{u}^{\left( s\right) }}{\lambda
_{t}^{\left( s\right) }}\frac{a}{p}\right) \\
&=&\lambda _{t}^{\left( s\right) }p\left( 1+\epsilon _{A}e^{i\phi _{A}}\frac{%
\lambda _{u}^{\left( s\right) }}{\lambda _{t}^{\left( s\right) }}\right)
\end{eqnarray*}
and 
\[
A\left( B^{0}\rightarrow K_{1}^{0}\gamma \right) =\lambda _{t}^{\left(
s\right) }p 
\]
where $\epsilon _{A}e^{i\phi _{A}}\equiv a/p$, $\phi _{A}$ is the strong
interaction phase and it disappears in $\mathcal{O}\left( \alpha _{s}\right) 
$ in the chiral limit. Hence we will set it equal to zero in the further
calculation. Following the same lines as for the charged $B$ meson the ratio
of the branching ratios for charged to neutral $B$ meson decays can be
written as 
\begin{equation}
\frac{\mathcal{B}\left( B^{-}\rightarrow K_{1}^{-}\gamma \right) }{\mathcal{B%
}\left( B^{0}\rightarrow K_{1}^{0}\gamma \right) }\simeq \left| 1+\epsilon
_{A}e^{i\phi _{A}}\frac{V_{ub}V_{us}^{*}}{V_{tb}V_{ts}^{*}}\right| ^{2}
\label{annihi3}
\end{equation}
The estimates in the frame work of the light-cone QCD sum rules yields
typically \cite{Ali-Braun, Stoll}; $\epsilon _{A}=-0.35$ and $\epsilon
_{A}=0.046$ for the decays $B^{-}\rightarrow K_{1}^{-}\gamma $ and $%
B^{0}\rightarrow K_{1}^{0}\gamma $, respectively. Let's define 
\begin{equation}
\frac{V_{ub}V_{us}^{*}}{V_{tb}V_{ts}^{*}}=-\left| \frac{V_{ub}V_{us}^{*}}{%
V_{tb}V_{ts}^{*}}\right| e^{i\alpha }=F_{1}+iF_{2}  \label{annihi4}
\end{equation}
where $\alpha $ is the unitarity triangle phase.

We also recall that the operator basis in $\mathcal{H}_{eff}$ is larger than
what is shown in Eq. (\ref{2}) in which the operator multiplying the CKM
factor$V_{ub}V_{us}^{*}$ have been neglected. To calculate CP-asymmetry we
have to put them back. Doing this, and using the unitarity relation $%
V_{cb}V_{cs}^{*}=-V_{ub}V_{us}^{*}-V_{tb}V_{ts}^{*}$ the effective
Hamiltonian reads \cite{Asatrian} 
\begin{eqnarray}
\mathcal{H}_{eff} &=&-\frac{G_{F}}{\sqrt{2}}\left\{ 
\begin{array}{c}
V_{tb}V_{ts}^{*}[C_{7}(\mu )O_{7}(\mu )+C_{8}(\mu )O_{8}(\mu )+C_{1}(\mu
)O_{1}(\mu )+C_{2}(\mu )O_{2}(\mu )] \\ 
V_{ub}V_{us}^{*}[C_{1}(\mu )(O_{1u}(\mu )-O_{1}(\mu ))+C_{2}(\mu
)(O_{2u}(\mu )-O_{2}(\mu ))+\ldots ]
\end{array}
\right\} .  \nonumber \\
&&  \label{newhamiltonian}
\end{eqnarray}
In the above equation the ellipses denote the terms proportional to the
Wilson coefficients $C_{3}\ldots C_{6}$ and we have dropped them because
they are very small as compared to $C_{1}$ and $C_{2}$. The operators $O_{1u}
$ and $O_{2u}$ are defined as 
\begin{eqnarray*}
O_{1u}(\mu ) &=&\left( \bar{s}_{L}\gamma _{\mu }T^{a}u_{L}\right) \left( 
\bar{u}_{L}\gamma ^{\mu }T_{a}b_{L}\right)  \\
O_{2u}(\mu ) &=&\left( \bar{s}_{L}\gamma _{\mu }u_{L}\right) \left( \bar{u}%
_{L}\gamma ^{\mu }b_{L}\right) 
\end{eqnarray*}
The values of Wilson coefficients in Eq. (\ref{newhamiltonian}) are same as
we have already used in Eqs. (\ref{eq:A1tb-C7}-\ref{eq:A1tb-sp}). Thus by
including the annihilation contribution and also the effect of the operator $%
O_{1u}$ and $O_{2u}$, the branching ratio from Eq. (\ref{branching1}) can be
written as 
\begin{eqnarray}
\mathcal{B}_{\mathrm{th}}(B^{\pm }\to K_{1}^{\pm }\gamma ) &=&\tau
_{B^{+}}\,\Gamma _{\mathrm{th}}(B^{\pm }\to K_{1}^{\pm }\gamma )  \nonumber
\\
&=&\tau _{B^{+}}\,\frac{G_{F}^{2}\alpha |V_{tb}V_{ts}^{*}|^{2}}{32\pi ^{4}}%
\,m_{b,\mathrm{pole}}^{2}\,M^{3}\,\left( 1-\frac{m_{K_{1}}^{2}}{M^{2}}%
\right) ^{3}\left[ \xi _{\perp }^{(K_{1})}(0)\right] ^{2}\,  \nonumber \\
&\times &\left\{ (C_{7}^{(0)\mathrm{eff}%
}+A_{R}^{(1)})^{2}+(F_{1}^{2}+F_{2}^{2})\,(A_{R}^{u}+L_{R}^{u})^{2}\right.  
\nonumber \\
&+&\left. 2F_{1}\,[C_{7}^{(0)\mathrm{eff}%
}(A_{R}^{u}+L_{R}^{u})+A_{R}^{(1)}L_{R}^{u}]\mp 2F_{2}\,[C_{7}^{(0)\mathrm{%
eff}}A_{I}^{u}-A_{I}^{(1)}L_{R}^{u}]\right\} ~,  \nonumber \\
&&  \label{ndecayrate}
\end{eqnarray}
where $L_{R}^{u}=\epsilon _{A}C_{7}^{(0)\mathrm{eff}}$ and the subscripts $R$
and $I$ denote the real and imaginary parts of the quantities involved. $%
A^{(1)}$ is same as defined in Eq. (\ref{eq:A1tb}) and $A^{u}$ corresponds
to the contribution from $O_{1u}$ and $O_{2u}$ which can be written as 
\begin{equation}
A^{u}(\mu )=\frac{\alpha _{s}(\mu )}{4\pi }\,C_{2}^{(0)}(\mu )\,\left[
r_{2}(z)-r_{2}(0)\right] -\frac{\alpha _{s}(\mu _{\mathrm{sp}})}{18\pi }%
\,C_{2}^{(0)}(\mu _{\mathrm{sp}})\,\frac{\Delta F_{\perp }^{(K_{1})}(\mu _{%
\mathrm{sp}})}{\xi _{\perp }^{(K_{1})}(0)}\,\frac{h^{(K_{1})}(z,\mu _{%
\mathrm{sp}})}{\left\langle \bar{u}^{-1}\right\rangle _{\perp
}^{(K_{1})}(\mu _{\mathrm{sp}})}.  \label{upenguin}
\end{equation}
We now proceed to calculate numerically the branching ratios for the decay $%
B^{+}\to K_{1}^{+}\gamma $. Using the value of CKM\ elements from \cite
{Eidelman}, the values of $A^{(1)}(\mu )$ from \cite{Jamil} and the value of 
$C_{2}^{(0)}(\mu )\,$from\cite{Greub:1996tg,Chetyrkin:1997vx}, the branching
ratio is plotted with unitarity triangle phase $\alpha $ as shown in Fig.
[7].

One can easily see that varying the value of $\alpha $ in the range $%
77^{0}\leq \alpha \leq 113^{0}$ with $\alpha =93^{0}$ as the central value,
there is a slight change in the value of branching ratio for the decay $%
\,B\rightarrow K_{1}(1270)\gamma $ leaving the value of the form factor to
be unchanged in this range as shown in the Fig. [8]. We also note that the
region of $\alpha $ where the branching ratio is effected is not allowed by
the CKM unitarity constraints within the SM which typically yields $%
77^{0}\leq \alpha \leq 113^{0}$.

We now compute the leading order CP-asymmetry $\mathcal{A}_{CP}\left(
K_{1}^{\pm }\gamma \right) $ for the decay $B^{\pm }\rightarrow K_{1}^{\pm
}\gamma $. The CP-asymmetry arises from the interference of the penguin
operator $O_{7}$ and the four-quark operator $O_{2}$\cite{Soares,Greub}. The
direct CP-asymmetry in the $B^{\pm }\rightarrow K_{1}^{\pm }\gamma $ is: 
\begin{eqnarray}
\mathcal{A}_{CP}\left( K_{1}^{\pm }\gamma \right)  &=&\frac{\mathcal{B}%
\left( B^{-}\rightarrow K_{1}^{-}\gamma \right) -\mathcal{B}\left(
B^{+}\rightarrow K_{1}^{+}\gamma \right) }{\mathcal{B}\left(
B^{-}\rightarrow K_{1}^{-}\gamma \right) +\mathcal{B}\left( B^{+}\rightarrow
K_{1}^{+}\gamma \right) }  \nonumber \\
&=&\frac{2F_{2}\left( A_{I}^{u}-\epsilon _{A}A_{I}^{\left( 1\right) }\right) 
}{C_{7}^{(0)\mathrm{eff}}\left( 1+2\epsilon _{A}\left[ F_{1}+\frac{1}{2}%
\epsilon _{A}\left( F_{1}^{2}+F_{2}^{2}\right) \right] \right) }
\label{cpasymmetry1}
\end{eqnarray}
The dependence of CP-asymmetry on different parameters involved is shown in
Fig. [9] and Fig. [10]. In Fig. [9] we have plotted the CP-asymmetry vs the
unitarity triangle phase $\alpha $. It is seen that in the SM favored
interval of $\alpha $, $77^{0}\leq \alpha \leq 113^{0}$, the CP-asymmetry
increases and reaches to its maximum value (on negative side like $K^{*}$)
which is $-0.75\%$ which reduces to the value $-0.45\%$ if one includes the
hard spectator corrections in addition to the vertex corrections and
annihilation contributions.

Fig. [10] shows the plot of $\mathcal{A}_{CP}\left( K_{1}^{\pm }\gamma
\right) $ with $\alpha $ at different values of the scale $\mu $. It is very
clear that the CP-asymmetry has the marked dependence on the scale $\mu $.
The maximum value of CP-asymmetry decreases from $-0.8\%$ to $-0.3\%$ in the
interval $m_{b\text{,pole}}/2\leq \mu \leq 2m_{b\text{,pole}}$. A similar
discussion for $B\rightarrow \rho \gamma $ is given in \cite{Ali}.

In conclusion, we have incorporated the effect of annihilation and $u$-quark
penguin contributions on the branching ratio for the decay $B\rightarrow
K_{1}(1270)\gamma $. It is shown that the value of LEET form factor remains
the same even with inclusion of these annihilation contributions for the
value of unitarity triangle phase $\alpha $ favored by Standard Model. Then
CP-asymmetry $\mathcal{A}_{CP}\left( K_{1}^{\pm }\gamma \right) $ for $%
\,B\rightarrow K_{1}(1270)\gamma $ has also been calculated. The
CP-asymmetry received contribution from the hard-spectator corrections which
tend to decrease its value estimated from the vertex corrections alone.
Unfortunately, the predicted value of CP-asymmetry is sensitive to the
choice of scale as well as to the quark mass ratio. The typical value of
CP-asymmetry lies around $-0.5\%$ which is almost same as for $B$ to $K^{*}$
decays.

\textbf{Acknowledgments}

The author would like to thank Prof. Fayyazuddin, Prof. Riazuddin and Prof.
Faheem Hussain for valuable discussion. This work was supported by a grant
from Higher Education Commission of Pakistan.

\begin{quote}
\textbf{Figure Captions}

a): Leading order contribution by operator $O_{7}$.

1): Feynman diagram contributing to the spectator corrections involving the $%
O_{7}$ operator in the decay $B\rightarrow K_{1}\gamma $. The curly (dashed)
line here and subsequent figures represents a gluon (photon).

2): Feynman diagram contributing to the spectator corrections involving the $%
O_{8}$ operator in the decay $B\rightarrow K_{1}\gamma $.

Row a: Photon is emitted from the flavor-changing line

Row b: Photon radiation off the spectator quark line.

3): Feynman diagram contributing to the spectator corrections involving the $%
O_{2}$ operator in the decay $B\rightarrow K_{1}\gamma $.

Row a: Photon is emitted from the flavor-changing line

Row b: Photon radiation off the spectator quark line.

4): Feynman diagram contributing to the spectator corrections involving the $%
O_{2}$ operator for the case when both the photon and virtual gluon are
emitted from the internal (loop) quark line.

5): Feynman diagram contributing to the spectator corrections involving the $%
O_{2}$ operator for the case when only the photon is emitted from the
internal (loop) quark line in the $bs\gamma $ vertex.

6): Branching ratio for $B\rightarrow K_{1}\gamma $ decay vs unitarity
triangle phase $\alpha $.

7): Branching ratio for $B\rightarrow K_{1}\gamma $ decay vs LEET form
factor for fixed value of $\alpha =93^{0}$.

8): CP-asymmetry ($\mathcal{A}_{CP}\%$) vs the unitarity triangle phase $%
\alpha $; dashed line shows the value without hard spectator correction and
solid line shows the value with hard spectator correction.

9): CP-asymmetry ($\mathcal{A}_{CP}\%$) vs the unitarity triangle phase $%
\alpha $ for different value of the scale $\mu $; dashed line shows the
value at $m_{b,\,pole}/2$; solid line shows the value at $m_{b,\,pole}$ and
the dotted line shows at $2m_{b,\,pole}$.
\end{quote}

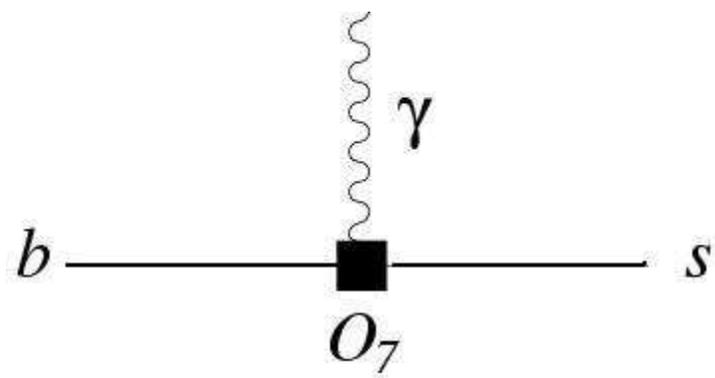
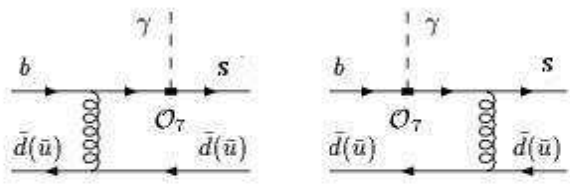
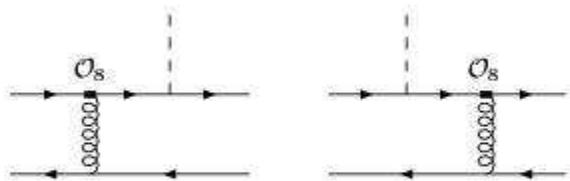
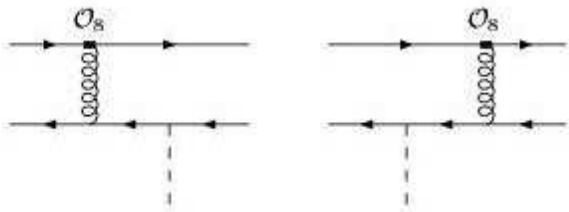
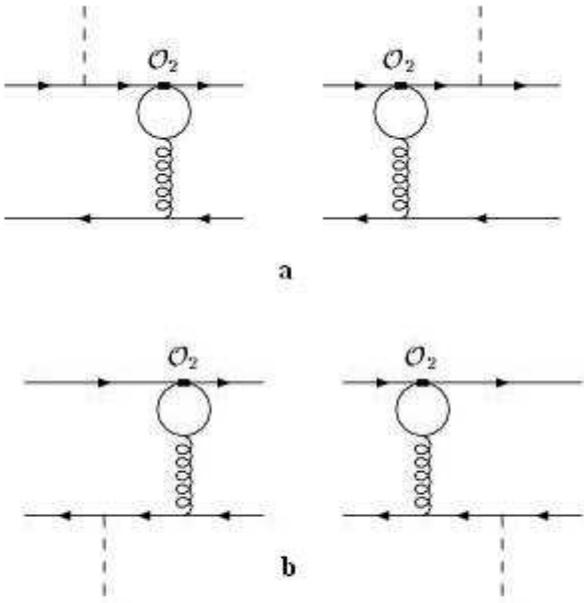
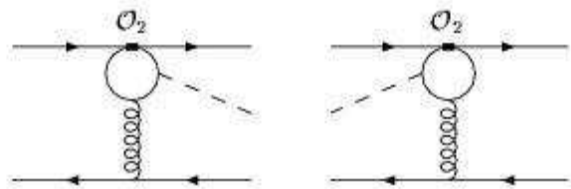
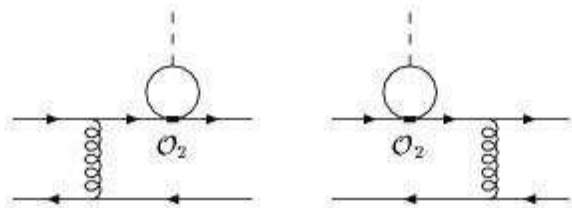
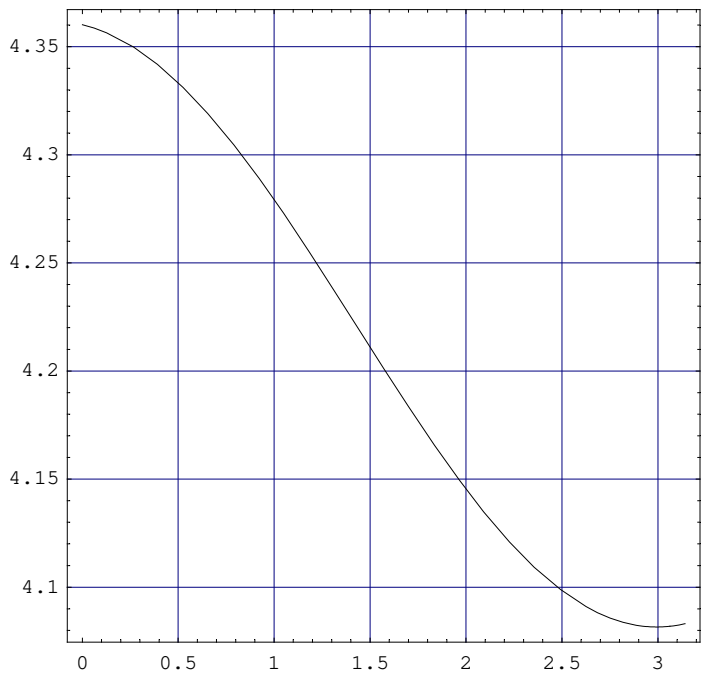
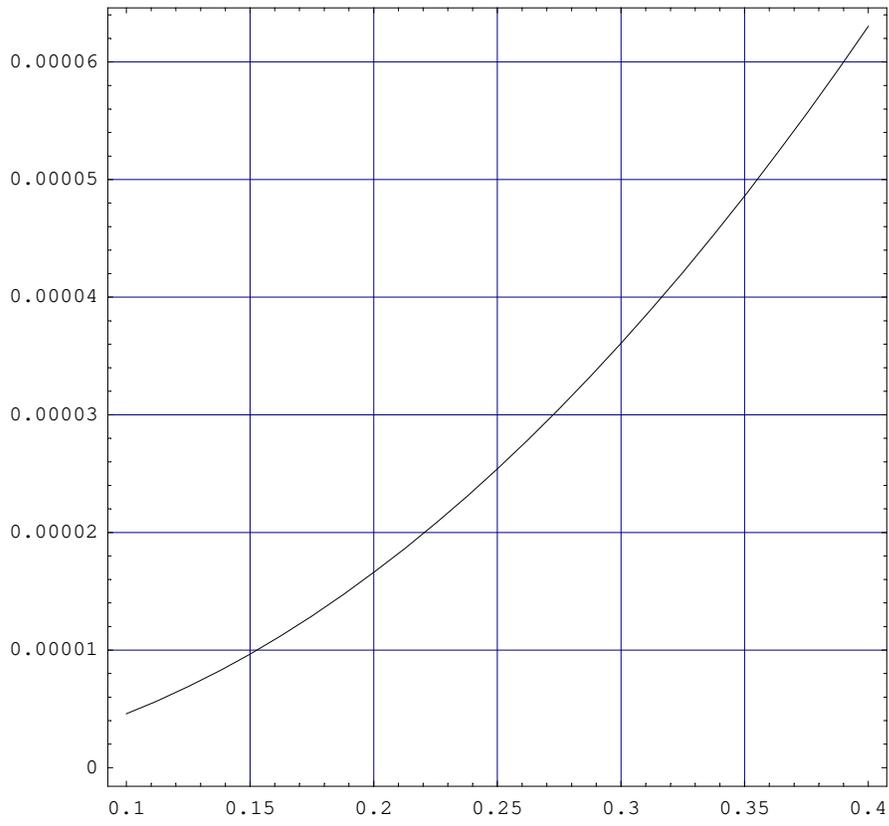
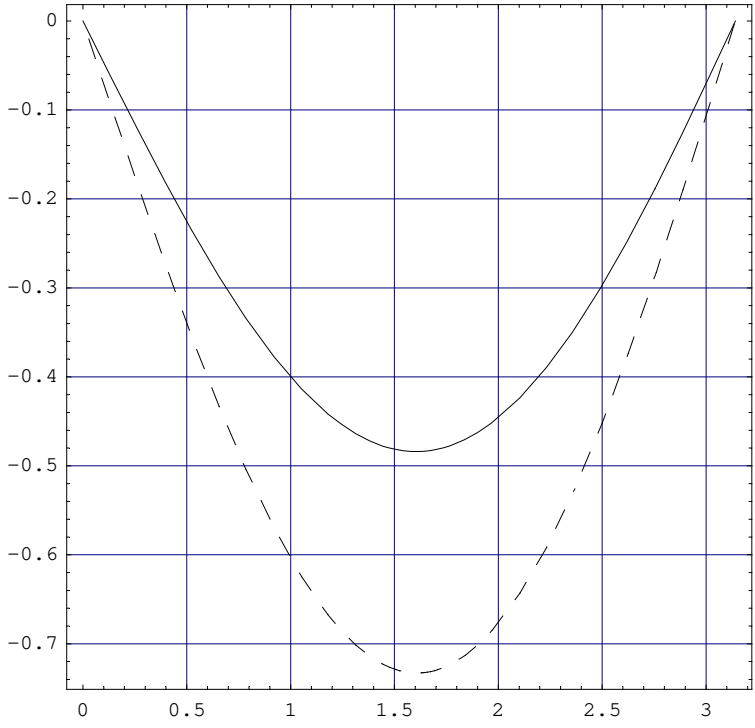
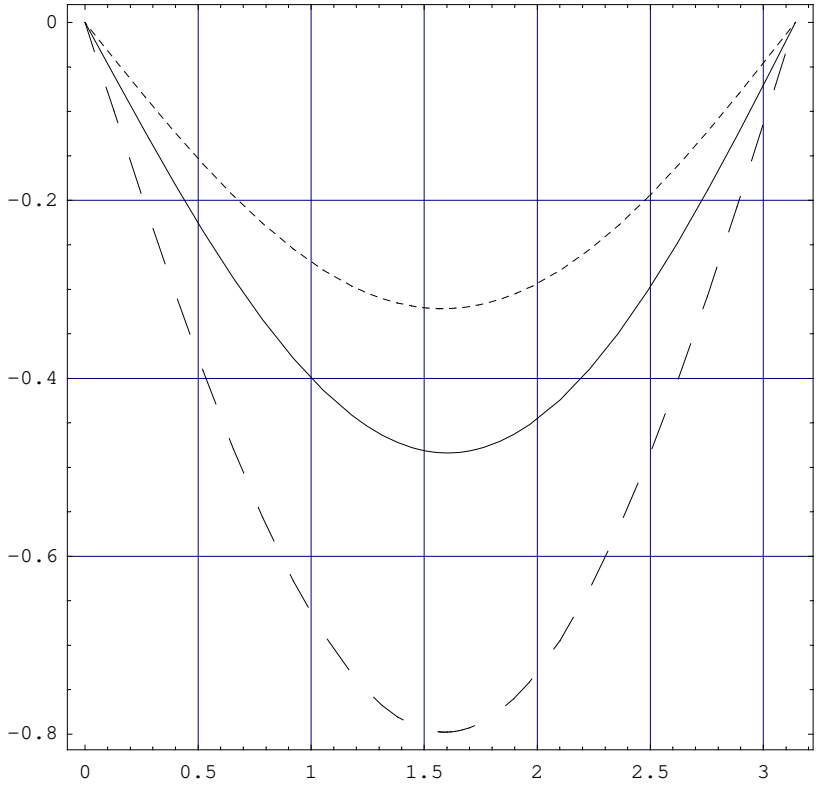
\end{document}